\documentclass[11pt,letter]{article}
  \usepackage{graphicx,subfigure}
  \usepackage{epstopdf}
  \usepackage{color}
    \usepackage{xcolor}
  \usepackage{epsfig,amssymb,amsmath}
  mm\evensidemargin=-5true mm \topmargin=-15true mm

\usepackage[
      colorlinks=true,
      linkcolor=blue,
      urlcolor=blue,
      filecolor=blue,
      citecolor=red,
      pdfstartview=FitV,
      pdftitle={},
      pdfauthor={},
      pdfsubject={},
      pdfkeywords={},
      pdfpagemode=None,
      bookmarksopen=true
]{hyperref}

 \newtheorem{prop}{Proposition}
\newtheorem{lemma}[prop]{Lemma}

\definecolor{AV}{rgb}{0.65,0.0,0}
\definecolor{geo}{rgb}{0,0.0,0.65}
\definecolor{GC}{rgb}{0,0.0,0.65}

  \newcommand{\rom}[1]{\mathrm{#1}}

  \newcommand{\be}{\begin{equation}}
  \newcommand{\ee}{\end{equation}}
  \newcommand{\bea}{\begin{eqnarray}}
  \newcommand{\eea}{\end{eqnarray}}
  
    \newcommand{\eps}{\epsilon}

  \newcommand{\s}{\sigma}
  
  \newcommand{\nn}{\nonumber}
\newcommand{\curl}{\mbox{curl}\,}
\newcommand{\sech}{\,\text{sech}\,}

\newcommand{\beq}{\begin{equation}}
\newcommand{\eeq}{\end{equation}}
\newcommand{\beqs}{\begin{eqnarray}}
\newcommand{\eeqs}{\end{eqnarray}}

\newcommand{\DD}{{\mathcal{D}}}

\begin{document}

\begin{flushright}
ULB-TH/11-08 \\
\today
\end{flushright}

\begin{center}
{\Large \bf
On Asymptotic Flatness and Lorentz Charges }
 \vspace*{0.5cm}\\
 {Geoffrey Comp\`ere$^{\flat}$, Fran\c cois Dehouck$^{\natural}$ and Amitabh Virmani$^{\natural}$}
\end{center}
 \vspace*{0.1cm}

 \begin{center}
 $^{\flat}${\it
 KdV Institute for Mathematics and Instituut voor Theoretische Fysica \\
Universiteit van Amsterdam, The Netherlands
 }
 \vspace*{0.5cm}\\
 $^{\natural}${\it
 Physique Th\'eorique et Math\'ematique,  \\ Universit\'e Libre de
     Bruxelles and International Solvay Institutes, Bruxelles,  Belgium
 }
 \vspace*{0.5cm}\\
{\tt gcompere@uva.nl} $\quad$  {\tt fdehouck@ulb.ac.be} $\quad$  {\tt avirmani@ulb.ac.be}

 \end{center}

\vspace{0.5cm}

\begin{center} \textit{ A la m\' emoire de notre ami et professeur Laurent Houart.} \end{center}

\vspace{0.5cm}

\begin{abstract}
In this paper we establish two results concerning four-dimensional asymptotically flat spacetimes at spatial infinity. First, we show that the six conserved Lorentz charges are encoded in two unique, distinct, but mutually dual symmetric divergence free tensors that we construct from the equations of motion. Second, we show that integrability of Einstein's equations in the asymptotic expansion is sufficient to establish the equivalence between counter-term charges defined from the variational principle and charges defined by Ashtekar and Hansen. These results clarify earlier constructions of conserved charges in the hyperboloid representation of spatial infinity. In showing this, parity condition on the mass aspect is not needed. Along the way in establishing these results, we prove two lemmae on tensor fields on three dimensional de Sitter spacetime stated by Ashtekar-Hansen and Beig-Schmidt and state and prove three additional lemmae.
\vspace{12pt}

\noindent
PACS: 04.20.-q 	Classical general relativity, 04.20.Ha Asymptotic structure
\vspace{12pt}
 \end{abstract}

\newpage

\tableofcontents

\section{Introduction and Summary of Results}

The study of asymptotically flat spacetimes of vacuum Einstein gravity at spatial infinity has a long history. Nevertheless, the topic has  constantly been evolving through the years,  see e.g. \cite{Dirac:1958jc,Deser:1960zzc,Arnowitt:1961zz,Arnowitt:1962aa, Regge:1974zd,Geroch:1977aa,Ashtekar:1978zz} for a relevant sample of classic works before the eighties, \cite{Witten:1981mf,Abbott:1981ff,Beig:1987aa,Ashtekar:1991vb,Wald:1993nt,Iyer:1994ys,Wald:1999wa,Szabados:2003yn,Mann:2005yr} for a sample of works in the last thirty years and \cite{Wald:1993nt,Iyer:1994ys,Wald:1999wa,Barnich:2001jy,Deser:2002rt,Deser:2002jk,Deser:2003vh,
Bouchareb:2007yx,Deser:2007vs,Barnich:2007bf} for more recent works addressing higher curvature terms in the action. Even nowadays, some subtleties in definitions of conserved charges at spatial infinity or in hypotheses underlying validity of variational principles are not completely settled. There are at least two motivations to further investigate  asymptotically flat spacetimes at spatial infinity.

First, there has been recent progress in the realization of gravitational electric-magnetic duality.  In the linear theory around flat space, the spin-2 Fronsdal action is invariant under a $SO(2)$ rotation generalizing the electric-magnetic duality of electromagnetism to linearized Einstein gravity \cite{Henneaux:2004jw, Barnich:2008ts} (see also \cite{Argurio:2008zt, Henneaux:2007ej, Hull:2001iu} and references therein). It has been shown that this duality of the linear theory cannot be extended perturbatively to the 3-vertex in Einstein gravity \cite{Deser:2005sz} using a proof similar to the one showing that electric-magnetic duality of free Maxwell theory cannot be extended to Yang-Mills theory \cite{Deser:1976iy}. It is also well understood that in Einstein gravity with a Killing symmetry, the mass and NUT charge of Taub-NUT spacetime can be rotated into one another upon acting with the $SO(2)$ generator of the Ehlers group obtained after dimensional reduction. More generally in supergravity theories, the Ehlers gravitational duality can be understood as an element of the U-dualities. These dualities were first described in dimensionally reduced theories assuming the presence of Killing vectors \cite{Julia:1980aa}. Quite remarkably, recent progress indicates that U-dualities might be a symmetry of supergravity theories without assuming the existence of Killing vectors \cite{Hull:2009mi,Berman:2010is,Hillmann:2009ci}. Furthermore, it has been shown that Taub-NUT spacetime is asymptotically flat in the sense of Regge and Teitelboim \cite{Bunster:2006rt}. In the light of these results, it is tempting to speculate that gravitational electric-magnetic duality with certain appropriately defined asymptotically flat boundary conditions might be realized in a non-perturbative or perhaps in a non-local sense so that it is not amenable to the treatment of \cite{Deser:2005sz}. We shall not pursue this direction further in this paper but we take this possibility as a motivation.

A second motivation for revising and exploring previous constructions of asymptotically flat spacetimes is the still elusive nature of holography in flat space, see e.g. \cite{Susskind:1998vk, Polchinski:1999yd, deBoer:2003vf, Arcioni:2003xx, Alvarez:2004hga, Marolf:2006bk, Barbon:2008ut, Barnich:2010eb, Li:2010dr,Compere:2011dx} and references therein for attempts in this direction. Clear indications of holography were found in the asymptotic structure of anti-de Sitter spacetime \cite{Brown:1986nw} before the AdS/CFT correspondence was formulated \cite{Maldacena:1997re,Witten:1998qj}. A universal object that one can derive from a given bulk configuration is the conserved stress-energy tensor expressed as a variational derivative of the renormalized action with respect to the boundary metric \cite{Henningson:1998gx,Balasubramanian:1999re,deHaro:2000xn,Bianchi:2001kw}.  In the asymptotically flat context, although there are many subtleties, a satisfactory stress-tensor construction is now available \cite{Mann:2005yr, Mann:2006bd, Mann:2008ay} for hyperbolic spatial and temporal cutoffs \footnote{Recall that boundary $\partial \mathcal M$ of an asymptotically flat spacetime $\mathcal M$ is not uniquely defined by the bulk spacetime. It depends on the choice of the limiting procedure used to define, say, the Gibbons-Hawking term. We refer the reader to \cite{Mann:2005yr} for a detailed discussion of these issues. In this work we exclusively work with hyperbolic spatial and hyperbolic temporal cutoffs.}. The stress-tensor is related in certain precise sense to the electric and magnetic parts of the Weyl tensor as shown in \cite{Mann:2006bd, Mann:2008ay}. On general grounds it was argued in \cite{Mann:2005yr}, following \cite{HIM}, that Poincar\'e generators defined by the boundary stress-tensor should agree with other approaches \cite{Arnowitt:1962aa,Regge:1974zd,Geroch:1977aa,Ashtekar:1978zz,Abbott:1981ff,Beig:1987aa,Ashtekar:1991vb}. However, the details needed to show an agreement \cite{Mann:2006bd, Mann:2008ay} to Ashtekar-Hansen and related approaches are quite non-trivial, to an extent that the underpinnings of why these calculations work are not clearly understood. One aim of this work is to clarify this.

We address two problems in the framework of Beig-Schmidt expansion \cite{BS,B} near spatial infinity in the standard definition of asymptotically flat spacetimes in four dimensions without NUT charges. First, we further elaborate on the discussion of \cite{Mann:2005yr, Mann:2006bd, Mann:2008ay} on how the six Lorentz charges are encoded at the boundary in the form of conserved tensors. We point out that there is not a unique conserved-tensor encoding these charges. Rather,  a \emph{pair} of uniquely defined conserved tensors  $V_{ab}$ and $W_{ab}$ can be associated with Lorentz charges. Charges constructed using either $V_{ab}$ or $W_{ab}$ are well defined in the sense that adding additional symmetric and divergence-free tensors constructed from non-linear combinations of the first order field does not change the values of these charges.  The fact that two conserved tensors appear at second order might be surprising, but stems from the remarkable properties of Killing vectors and tensors on three-dimensional de Sitter space. In short, rotations and Lorentz boosts are curl of each other,
\bea
\DD \times \xi_\rom{rot} = -2 \xi_\rom{boost},\qquad \DD \times \xi_\rom{boost} = +2\xi_\rom{rot},\label{dualxi}
\eea
and the conserved tensors $V_{ab}$ and $W_{ab}$ are also curl of each other up to conserved tensors that do not contribute to charges. It then follows that  Lorentz charges can be expressed in two equivalent ways using either $V_{ab}$ or $W_{ab}$, as developed in the main text. This observation was also made in \cite{Mann:2006bd, Mann:2008ay}, where
$V_{ab}$ and $W_{ab}$ are directly related to the second order electric and magnetic parts of the Weyl tensor. In this work we emphasize that these are general properties of regular conserved tensors on three-dimensional de Sitter space and the analysis of \cite{Mann:2006bd, Mann:2008ay} can be regarded as a special case of the discussion developed in the main text.

The second issue we address is the precise boundary condition on the mass aspect  (field $\sigma$ in the Beig-Schmidt expansion) required to construct conserved charges of asymptotically flat gravity.  In this regard we first recall that there is a slight tension in the literature. To show that the counter-term prescription of \cite{Mann:2005yr} gives a well defined variational principle for asymptotically flat gravity, parity condition on the mass aspect is not required. However, to show \cite{Mann:2006bd, Mann:2008ay} that the charges defined in \cite{Mann:2005yr} are equivalent to the Ashtekar-Hansen charges \cite{Ashtekar:1978zz, AshRev}, the parity condition on the mass aspect was explicitly assumed.  This creates some tension: one would expect that a well defined action principle leads to a finite and conserved symplectic structure and finite and conserved charges without requiring any extra boundary condition. In this work, we solve partially this tension. We show that the parity condition on the mass aspect is not required in order to establish the equivalence between the counter-term charges and the Ashtekar-Hansen charges --- integrability of Einstein's equations in the metric expansion is sufficient. We differ to further work \cite{Compere:2011bb} the resolution of the fact that the symplectic structure is finite only when parity conditions are imposed \cite{ABR} while the action is finite for general values of the mass aspect without reference to any parity property \cite{Mann:2005yr}. We also note that conserved charges associated with Lorentz boosts are not linear functionals of the mass aspect. This result is in sharp contrast with the linearity in the boundary fields of standard ADM conserved charges \cite{Deser:1960zzc,Arnowitt:1961zz,Arnowitt:1962aa}. Linearity is only recovered once the mass aspect is restricted to be parity even. This creates a new tension between definitions of conserved charges considered in this paper and the ones defined in Hamiltonian formalism. We will also differ the issue of resolving this tension to a forthcoming publication \cite{Compere:2011bb}.

The so-called logarithmic translation ambiguities \cite{logB} in the definition of asymptotic fields are usually fixed by the parity condition on the mass aspect \cite{AshLog}. We choose to fix these ambiguities by imposing that the four lowest harmonics on $dS_3$ of the mass aspect are parity even. This condition is weaker than imposing that the entire mass aspect is parity even. We also observe that the parity on the four lowest harmonics of $\s$ on $dS_3$ is all that is needed for the arguments of \cite{AshNew} to go through.

Along the way in establishing our arguments, we use harmonic decomposition of various tensor fields on three dimensional hyperboloid to provide proofs of two lemmae due to Ashtekar and Hansen \cite{Ashtekar:1978zz, AshRev} and Beig and Schmidt \cite{BS}. To the best of our knowledge, proofs of these  quite non-trivial lemmae have not been reported in the literature before.  We, thereby, fill this gap.  We also provide a characterization of tensor fields appearing at second and higher orders in the asymptotic expansion in three additional lemmae.  All lemmae are stated and proven in Appendix \ref{proptensors}. Appendix \ref{vecKilling} collects useful properties of Killing vectors and associated charges on $dS_3$. We emphasize that Appendices \ref{proptensors} and \ref{vecKilling} are an essential part of this paper. The lemmae and properties of Killing vectors proven there are used in a number of ways throughout the main text.

The rest of the paper is organized as follows. Section \ref{prelim} provides relevant definitions and review and our precise boundary conditions. In Section \ref{oneold}, we present asymptotic equations of motion in two equivalent forms and discuss Beig's integrability conditions. In Section \ref{one} we present our general construction of conserved Lorentz charges from the equations of motion while in Section \ref{two} we show the equivalence between Lorentz counter-term charges and Ashtekar-Hansen charges. A classification of tensor structures useful for the main arguments is relegated to Appendix \ref{classSTD}.

\section{Asymptotic Flatness, Variational Principle, and Stress Tensor}
\label{prelim}
In this section we provide relevant definitions and review. We introduce asymptotic flatness based on \cite{BS, B} and discuss diffeomorphisms preserving the asymptotic expansion. We obtain the Poincar\'e group as asymptotic symmetry group after fixing suitably the asymptotic frame and boundary conditions. We then review the variational principle of \cite{Mann:2005yr} and the associated boundary stress tensor \cite{Mann:2006bd, Mann:2008ay}. We exclusively work with a coordinate based definition of asymptotically flat spacetimes. The results we obtain in this paper can be readily translated to geometric language of Ashtekar-Hansen \cite{Ashtekar:1978zz, AshRev} or that of Ashtekar-Romano \cite{Ashtekar:1991vb}.

We define asymptotically flat spacetimes as spacetimes admitting an asymptotic expansion, usually referred to as the Beig-Schmidt expansion, at spatial infinity as
\be
\label{metric}
ds^2 = \left( 1 + \frac{\s}{\rho} \right)^2 d \rho^2 + \rho^2\left( h^{(0)}_{ab} + \frac{h^{(1)}_{ab}}{\rho} + \frac{h^{(2)}_{ab}}{\rho^2} + o(\rho^{-2}) \right) dx^a dx^b,
\ee
where $h^{(0)}_{ab}dx^ dx^b$ is the metric on the unit hyperboloid or, equivalently, on three-dimensional de Sitter space\footnote{The coordinates ranges are $\tau \in \mathbb R$, $\phi \in [0,2\pi)$, $\theta \in (0,\pi)$. We will denote $\DD_a$ as the covariant derivative on the unit hyperboloid. The radial coordinate $\rho$ is associated to some asymptotically Minkowski coordinates $x^\mu$ via $\rho^2 = \eta_{\mu \nu} x^\mu x^\nu$. The symbol $o(\rho^{-p})$ is defined such that $\rho^p o(\rho^{-p}) =0$ as $\rho \to \infty$ for fixed values of the angular coordinates $(\tau, \theta, \phi)$ of the hyperboloid.} $dS_3$,
\be
ds^2_{(0)} \equiv h^{(0)}_{ab}dx^a dx^b = -d\tau^2 + \cosh^2\tau (d\theta^2 + \sin^2\theta d\phi^2).\label{h0}
\ee
The fields $\sigma$, $h^{(1)}_{ab},h^{(2)}_{ab}$, etc are assumed to be smooth functions on the unit hyperboloid. We will use $h_{ab}$ in what follows to denote the complete induced metric on a constant $\rho$ slice for some large value of $\rho$.

The set of diffeomorphisms  preserving the Beig-Schmidt form \eqref{metric} has been analyzed by a number of authors over the years \cite{Ashtekar:1978zz, BS, B, Ashtekar:1991vb}, see \cite{AshNew} for a concise review. In an asymptotically cartesian coordinate system with $\rho^2 = \eta_{\mu \nu} x^{\mu} x^{\nu}$, diffeomorphisms
\be
\bar{x}^\mu = L_{\nu}^{\mu}x^\nu + T^\mu + S^\mu(x^a)  + o(\rho^0), \label{diffeos}
\ee
preserve the form of the metric \eqref{metric}. The transformations generated by the constants $L_{\nu}^{\mu}$ and $T^\mu$ form the Poincar\'e group, while transformations generated by angle dependent translations $S^\mu(x^a)$ are the so-called supertranslations.

In conventional treatments of asymptotic spacetimes, one requires that all asymptotic symmetries are associated with conserved and well-defined charges. Since supertranslations depend arbitrarily on the angular coordinates, the associated charges are in general not conserved. In fact, the philosophy that a large body of work on asymptotic flatness at spatial infinity has taken is to strengthen the boundary conditions so that the freedom of performing supertranslations is eliminated. Indeed, this can be achieved \cite{Ashtekar:1978zz, B, Ashtekar:1991vb, ABR, AshNew}, for instance, by demanding the leading order asymptotic Weyl curvature to be purely electric.  This condition removes altogether the possibility of NUT charges. This condition is fullfilled by choosing $h^{(1)}_{ab} = - 2 \sigma h^{(0)}_{ab}$.

There is an additional boundary condition that needs to be imposed. This has to do with the so-called logarithmic translations.
Logarithmic translations in the hyperbolic coordinates are written as
\be
\rho = \bar{\rho}  + \log \bar{\rho} H(\bar{x^a}) - H(\bar{x}^a) + o(\rho^0),  \qquad
x^a = \bar{x}^a + \frac{\log \bar{\rho}}{\bar{\rho}}H^a(\bar{x}^a)  + o(\rho^{-1}),
\ee
where $H$ obeys $\DD_a \DD_b H + H h^{(0)}_{ab} = 0$. Under a general log-translation, $\sigma$ and $h^{(1)}_{ab}$ transform as
\be
\sigma \rightarrow  \sigma + H, \qquad
h^{(1)}_{ab} \rightarrow   h^{(1)}_{ab} -2 H h_{ab}^{(0)}. \label{log1}
\ee
Not only these fields transform, but also an unwanted logarithmic term
\be
-\log \rho \: \DD_c (E^{(1)}_{ab} H^c) dx^a dx^b,
\ee is generated in the transformed metric. Since such a logarithmic term is not allowed in the metric \eqref{metric} to start with,  generically logarithmic translations are not asymptotic symmetries of our notion of asymptotic flatness. However, there is an important exception. If for certain choices of $H$ it happens that $\DD_c (E^{(1)}_{ab} H^c)=0$, then the unwanted log term in the metric is not generated. These special logarithmic translations are then allowed by our ansatz. This causes a number of problems, the most obvious one is that the set of asymptotic diffeomorphisms preserving the form \eqref{metric} are not just Poincar\'e. To deal with these problems, we simply demand that $\sigma$ does not contain in the harmonic decomposition on $dS_3$, the four solutions of the equation $\DD_a \DD_b H + H h^{(0)}_{ab} = 0$. In other words, we demand in the harmonic decomposition of $\sigma$ on $dS_3$ the four lowest components $l=0,m=0$, $l=1, m=-1,0,1$ to be parity even, see Appendix \ref{proptensors} for further details. References \cite{AshLog, ABR, AshNew} demand $\sigma(x^a)$ to be entirely reflection symmetric $\sigma(\tau, \theta, \phi) = \sigma(-\tau, \pi-\theta, \phi+\pi)$ in order to eliminate these logarithmic translations. Here, we see that a milder condition is sufficient to fix this ambiguity uniquely. It is also important to choose boundary conditions such that the canonical (Hamiltonian or Lagrangian) Poincar\'e generators are finite and well-defined. Here, we also observe that the even parity condition on the four lowest harmonics of $\s$ on $dS_3$ is all that is needed for the arguments of \cite{AshNew} to go through. The Poincar\'e charges can then be defined in a first order formalism \cite{AshNew} with very similar boundary conditions.

Finally, the set of allowed configurations is also restricted by the requirement that Einstein's equations can be solved at all orders in the expansion. As proven in \cite{BS, B}, a necessary and sufficient condition when only negative powers of $\rho$ are allowed in the metric expansion \eqref{metric} is that the  following six charges\footnote{Our conventions are $\s_a = \DD_a \s$, $\s_{ab} = \DD_b \DD_a \s$, $\s_{abc} = \DD_c \DD_b \DD_a \s$ etc.}
\bea
\mathcal Q [\xi_{(0)}] \equiv \int_S d^2S\, \eps_{cd(a}\s^c \s^d_{\; \, b)} \xi^a_{(0)} n^b =0\label{intO}
\eea
vanish, where $\xi_{(0)}^a$ are the six Killing vectors on the hyperboloid, $S$ is a Cauchy surface in the unit hyperboloid, and $n^{a}$ is a unit timelike vector normal to $S$ in the unit hyperboloid. We will see in section \ref{sec:intCond} that one can express these integrability conditions in another interesting form.

The phase space of asymptotically flat spacetimes is now completely defined.  The asymptotic symmetry group is the Poincar\'e group. The conditions mentioned above are met for a very large class of asymptotically flat spacetimes, with the notable exception of spacetimes with NUT charges.

It was shown in \cite{Mann:2005yr} that a good variational principle for asymptotically flat gravity defined by the expansion \eqref{metric} is given by the action
\begin{equation}
\label{covaction} S = \frac{1}{16\pi G} \int_{\cal M} \sqrt{-g}\,R + \frac{1}{8\pi G} \int_{\partial {\cal M}} \sqrt{-h} \,(K - \hat K),
\end{equation}
where $\hat K := h^{ab} \hat K_{ab}$ and $\hat K_{ab}$ is defined to satisfy
\begin{equation}
\label{Khat}
{\cal R}_{ab} = \hat K_{ab} \hat K - \hat K_a{}^{c} \hat K_{cb},
\end{equation}
where ${\cal R}_{ab} $ is the Ricci tensor of the boundary metric $h_{ab}$.
This equation being quadratic in $\hat K_{ab}$ admits more than one solution for $\hat{K}_{ab}$. To cancel divergences we choose the solution that asymptotes to the extrinsic curvature of the boundary of Minkowski space as $\partial {\cal M}$ (constant $\rho$ slice for some large value of $\rho$) is taken to infinity. It is important for our purposes to note that this counter-term gives a well defined variational principle without requiring symmetry condition on the mass aspect $\sigma$. This counter-term makes the idea of background substraction \cite{GH, BY, HH} precise. Earlier steps towards such a construction were taken in \cite{Kraus:1999di, Mann:1999pc, deHaro:2000wj}. See \cite{Astefanesei} for other interesting results obtained using counter-term methods in asymptotically flat spacetimes.

Varying the action and imposing the equations of motion yields
\begin{equation}
\label{ActionVariation}
\delta S = \frac{1}{16 \pi G} \, \int_{\partial M} \sqrt{-h}\,\left(\pi_{ab} - \hat{\pi}_{ab} + \Delta_{ab} \right)\delta h^{ab},
\end{equation}
where $\hat{\pi}_{ab} := h_{ab}\,\hat{K} - \hat{K}_{ab}$ and $\Delta_{ab}$ are the remaining terms. The expression for $\Delta_{ab}$ is not needed here but can be found in  appendix A of \cite{Mann:2008ay}. The boundary stress tensor is defined as the functional derivative of the on-shell action with
respect to $h_{ab}$ \begin{equation}\label{StressTensor} T_{ab} := -\frac{2}{\sqrt{-h}}\,\frac{\delta S}{\delta h^{ab}} = -\frac{1}{8 \pi G}\,\left(\pi_{ab} - \hat{\pi}_{ab} + \Delta_{ab}
\right).
\end{equation}
The asymptotic expansion of the stress tensor is
 \be
T_{ab} = \frac{1}{8 \pi G} \left( T^{(1)}_{ab} + \frac{ T^{(2)}_{ab} + \Delta^{(2)}_{ab}}{\rho} + o(\rho^{-1}) \right),\label{stress1}
 \ee
where \cite{Mann:2005yr, Mann:2006bd, Mann:2008ay}
\bea
T^{(1)}_{ab} &= & E^{(1)}_{ab} = -\s_{ab} -\s h^{(0)}_{ab},  \\
T^{(2)}_{ab} &=& -h^{(2)}_{ab} + \left( \frac{5}{2} \,\s^2 + \s_c \s^c +
\frac{1}{2} \,\s_{cd} \s^{cd} \right) h^{(0)}_{ab}  - 2\,\s_a
\s_b - \s \s_{ab} - \s_{a}{}^c\s_{cb}  \\
&=& E_{ab}^{(2)} - \gamma_{ab}, \\
\gamma_{ab}&=& 2 \s \s_{ab} + \frac{5}{2} \s^2 h^0_{ab} + \s_a^c \s_{cb} - \frac{1}{2} \s_{cd} \s^{cd} h^0_{ab},  \\
\label{DeltaPi2}
\Delta^{(2)}_{ab} &=&
-\frac{1}{4} \left[ 9 \s_c \s^c h^{(0)}_{ab} -29 \s_a \s_b + 63\s\s_{ab} + 24 \s_{ap}\s_{b}{}^{p} -5 \s_{cd}\s^{cd}h^{(0)}_{ab} + 45 \s^2 h^{(0)}_{ab}\right. \nn \\
& &
\left. \quad -  3\s_{mnp}\s^{mnp}h^{(0)}_{ab} + 9 \s_{pq(a}\s^{pq}{}_{b)} -3 \s^{pq}\s_{pq(ab)} - 2\s^e\s_{e(ab)} \right]. \label{Delta}
\eea
Here, $E_{ab}^{(1)}$ and $E_{ab}^{(2)}$ are the first and second order terms in the expansion of the electric part of the Weyl tensor. We discuss the equivalence of  conserved charges built from the stress-tensor and the Ashtekar-Hansen charges in Section \ref{two} after having presented a general analysis of Lorentz charges in Section \ref{one}.

\section{Equations of Motion and Integrability}
\label{oneold}
In this section we show that  Einstein's equations at second order in the asymptotic expansion can be written in two mutually dual and equivalent forms, in a sense that we precise herebelow. We also present an original rewriting of Beig's integrability conditions \cite{B} as the vanishing of the six Noether charges associated with the mass aspect $\sigma$ considered as a free scalar field on $dS_3$.

Given the form of the metric (\ref{metric}), it is natural to decompose vacuum Einstein's equations using the outward-pointing unit normal $\rho^a$ to the
hyperboloid of constant $\rho$ and the projector $h_{ab} = g_{ab} -\rho_a \rho_b$.  This  provides us with Hamiltonian and momentum equations of motion (these equations contain time derivatives and therefore are not constraints) and equations of motion on the 3-dimensional hypersurface which respectively read \cite{BS}
\beqs\label{3equationssimp}
H &\equiv&  -\mathcal{L}_{\rho} K - K_{ab}K^{ab}-N^{-1} h^{ab} D_a D_b N=0, \nonumber \\
F_{a} &\equiv& D_{b} K^{b}_{\:\:a}-D_a K=0, \nonumber \\
F_{ab} &\equiv &   \mathcal{R}_{ab}- N^{-1} \partial_{\rho} K_{ab}-N^{-1} D_a D_b N-K K_{ab}+ 2 K_{a}^{\:\:c} K_{cb}=0,
\eeqs
where $D$ is the covariant derivative compatible with the full metric
$h_{ab}$ on the hyperboloid, $N = 1 + \sigma/ \rho$ is the lapse function,  $\mathcal{L}_{\rho}$ is the Lie derivative in the direction of $\rho^a$, and $\mathcal{L}_{\rho}K = \mathcal{L}_{\rho}(h^{ab}K_{ab})$.  In  equations (\ref{3equationssimp}) indices are raised and lowered with $h_{ab}$.  In the rest of this section
 indices will be raised and lowered with $h^{(0)}_{ab}$.

\subsection{Equations of motion at second order}

Beig and Schmidt \cite{BS} showed that at the zeroth order, the equations of motion require $h^{(0)}_{ab}$ to be locally the metric on the unit hyperboloid. At first order demanding $h^{(1)}_{ab} = - 2 \s h^{(0)}_{ab}$, they showed that Einstein equations are identically satisfied if
\begin{equation} \square \s + 3 \s =0. \label{sigma}
\end{equation}
At second order Beig \cite{B} obtained
\begin{eqnarray}
h^{(2)}&=&12 \sigma^2 + \sigma_c \sigma^c, \nonumber \\
\DD^b h^{(2)}_{ab}&=& \DD_a \left( \sigma_c \sigma^c +8 \sigma^2 \right),\label{eqh2} \\
(\Box - 2) h^{(2)}_{ab} &=& 6 \s_c \s^c h_{ab}^{(0)} + 8\s_a \s_b +14 \s \s_{ab} -18 \s^2 h_{ab}^{(0)}+2\s_{ac} \s^{c}_{\; b}+2\s_{abc}\s^c. \nonumber
\end{eqnarray}

In an attempt to present the material in a more pedagogical fashion, we now discuss the above equations in the linear case, when the quadratic terms in $\sigma$ on the r.h.s of equations \eqref{eqh2} are set to zero.
The non-linear equations will be considered at the end of the section.

Let us define
\bea
V_{ab} &\equiv & - h_{ab}^{(2)}, \label{defV0}\qquad W_{ab} \equiv  \curl h_{ab}^{(2)} \label{defW0},
\eea
where the curl operator is $(\curl T)_{ab} \equiv  \eps_{a}^{\;\, cd}D_{c} T_{db}$. The curl operator obeys remarkable properties. The curl of a symmetric tensor $T_{ab}$ satisfying $\DD^bT_{ab} = \DD_a T_{b}{}^{b}$ is symmetric: $(\curl T)_{[ab]}=0$. Moreover, the curl is the square root of the operator $\square - 3$ when acting on Symmetric, Divergence-free, and Traceless (SDT) tensors $T_{ab}$
\be
\curl(\curl(T))_{ab} = (\square - 3 )T_{ab}.
\ee
The latter property also implies that the square of the curl operator when acting on an SDT tensor $T_{ab}$ obeying $(\square - 2)T_{ab} = 0$ is minus the identity. This shows that this operator is invertible when acting on $T_{ab}$ satisfying $(\square - 2)T_{ab} = 0$. Finally, one has $[\square - 2, \curl] T_{ab} = 0$ for an SDT tensor $T_{ab}$.

With these properties in mind, we see that the above defined quantities enjoy the duality properties
\bea
W_{ab} = - \curl V_{ab},\qquad V_{ab} = \curl W_{ab}.
 \eea
Moreover, the linearized second order equations  \eqref{eqh2} can then be written in two equivalent ways
\begin{align}
V^a_{\;a} &= 0,& \DD^b V_{ab} &= 0,&  (\square - 2)V_{ab} &= 0,\label{eq:001a}
\end{align}
or
\begin{align}
W^a_{\;a} &= 0,&  \DD^b W_{ab} &= 0,&  (\square - 2)W_{ab} &= 0\label{eq:001b}.
\end{align}
The two sets of equations are related by the curl operator. Given a solution to one of these systems, one can reconstruct the metric using definitions \eqref{defV0}. We have thus shown the equivalence of linearized Einstein's equations at second order to any one of the above two systems of equations.

At this stage, the reader might find useful to have a more precise idea of the form of the solutions to the equations
\bea
T^a_a = \DD^b T_{ab} = (\square - 2)T_{ab} = 0.\label{eq:Taa2}
\eea
In Appendix \ref{proptensors}, we prove Lemma \ref{lemma8} stating that the general solution of \eqref{eq:Taa2} consists of two sets of three tensors related by the curl operator that capture the six charges $\int_S T_{ab} \xi_{(0)}^a n^b$ associated with Lorentz transformations, supplemented by higher harmonic tensors that do not contribute to the charges $\int_S T_{ab} \xi_{(0)}^a n^b$. Note that tensor fields that derive from a scalar potential, like the first order electric part of the Weyl tensor $E_{ab}^{(1)}= - \DD_a \DD_b \sigma - h^{(0)}_{ab}\sigma$, have vanishing curl,
\bea
T_{ab} = \DD_a \DD_b \Phi + h^{(0)}_{ab} \Phi \quad \Rightarrow \quad (\curl T)_{ab} = 0.
\eea
These tensors therefore obey $(\curl^2 \,T)_{ab} = (\square - 3)T_{ab} = 0$ as opposed to $(\square - 2)T_{ab} = 0.$ As a consequence, none of the lemmae used in the first order analysis, e.g., in \cite{BS, Ashtekar:1978zz}, capture tensor structures appearing at second order. See also Appendix \ref{proptensors} for proofs of the unproven lemmae in \cite{BS, Ashtekar:1978zz, AshRev}.

Let us now rewrite Beig's equations \eqref{eqh2} including non-linearities in a form similar to the mutually dual set of equations \eqref{eq:001a}-\eqref{eq:001b}.

A direct calculation shows that the following tensor
\be
E^{(2)}_{ab} - \s E^{(1)}_{ab} = - h^{(2)}_{ab} + 6 \s^2 h^{(0)}_{ab} + 2 \s_{ab} \s - 2 \s_a \s_b + \s^c \s_c h^{(0)}_{ab},
\ee
constructed out of the asymptotic expansion of the electric part of the Weyl tensor is an SDT tensor. Similarly, the following tensor constructed out of the asymptotic expansion of the magnetic part of the Weyl tensor is also an SDT tensor
\be
B^{(2)}_{ab} = \epsilon_{cda}D^{c}(h^{(2)}_{b}{}^{d} - 2 \s^2 \delta_{b}{}^{d}),
\ee
on account of the equations of motion. The equations of motion can be reformulated in terms of $B^{(2)}_{ab}$ as \cite{B}
\bea
B^{(2)}_{a}{}^{a} &=& 0, \\
D^{a}B^{(2)}_{ab} &=& 0,\\
(\Box - 2)B^{(2)}_{ab} &=& - 4 \epsilon_{cd(a}\s^cE^{(1)}_{b)}{}^{d}. \label{B23eq}
\eea
In terms of $E^{(2)}_{ab} - \s E^{(1)}_{ab}$ the same equations take the form
\bea
(E^{(2)} - \s E^{(1)})_{a}{}^{a} &=& 0, \\
D^{a}(E^{(2)}_{ab} - \s E^{(1)}_{ab}) &=& 0,\\
(\Box - 2)(E^{(2)}_{ab} - \s E^{(1)}_{ab}) &=&  \curl [ -4  \epsilon_{cd(a}\s^cE^{(1)}_{b)}{}^{d}]. \label{E23eq}
\eea
Furthermore, it can be easily checked that
\be
E^{(2)}_{ab} - \s E^{(1)}_{ab} = \curl  B^{(2)}_{ab}, \qquad - \curl (E^{(2)}_{ab} - \s E^{(1)}_{ab}) = B^{(2)}_{ab} + 4  \epsilon_{cd(a}\s^cE^{(1)}_{b)}{}^{d}.
\ee
Note that $E^{(2)}_{ab} - \s E^{(1)}_{ab}$ is linear in $h_{ab}^{(2)}$, and $B^{(2)}_{ab}$ is linear in its curl. Therefore, $ E^{(2)}_{ab} - \s E^{(1)}_{ab} $ and $B^{(2)}_{ab}$ precisely reduce to $V_{ab}$ and $W_{ab}$ in the linear approximation.

\subsection{Integrability conditions}
\label{sec:intCond}

Solutions of linearized equations are not always linearizations of solutions of non-linear equations. This phenomenon is well-known as a linearization instability \cite{Deser:1973zza,Moncrief:1975aa,Moncrief:1976un}. In the present case of Beig-Schmidt expansion, it has been shown \cite{B} that there are exactly six obstructions in reconstructing non-linear asymptotic solutions when only negative powers of $\rho$ are allowed in \eqref{metric}. The existence of these six obstructions can be seen as follows. Contracting equation \eqref{B23eq} with a Killing vector on $dS_3$, one can rewrite the l.h.s of the expression as
\be
(\Box-2)B^{(2)}_{ab}  \xi_{(0)}^{a} =2\DD^{a}\left( \xi_{(0)} ^{c}\DD_{[a}B^{(2)} _{b]c}+B^{(2)} _{c[a}\DD_{b]}\xi_{(0)} ^{c}\right),
\label{Beig30}
\ee
which is a total divergence and vanishes when integrated on a Cauchy surface $S$ on the unit hyperboloid. For consistency, we must require that the integrals on the sphere of the r.h.s of \eqref{B23eq} contracted with $\xi_{(0)}^{a}$ are also zero. These requirements are precisely Beig's integrability conditions \eqref{intO}\footnote{This argument only shows that these conditions are the necessary conditions. The arguments showing that these conditions are also sufficient to solve Einstein's equations to all orders in the expansion can be found in \cite{B}. Note that there is a minor typo in eq (43) of \cite{B}. The coefficient multiplying $\beta_a$ is $(n-1)^2$ instead of $(n+1)^2$.}.

We now present a new way  of looking at these integrability conditions. It is clear that the equation for the mass aspect $\sigma$ \eqref{sigma} can be derived from the free scalar Lagrangian $L^{(\sigma)}$
\bea
L^{(\sigma)} = \sqrt{-h_{(0)}} \Big( -\frac{1}{2} \partial_a \s \partial^a \s +\frac{3}{2}\s^2 \Big),\label{Ls}
\eea
with mass $m^2 = -3$ on three-dimensional de Sitter space. Now, it is interesting to note that Beig's integrability conditions are precisely the conditions that all six Noether charges derived from this Lagrangian vanish. Indeed, one has $ \eps_{cd(a}\s^c \s^d_{\;\, b)}  = -  (\curl \kappa)_{(ab)} $ where
\bea
\kappa_{ab} &=& - \frac{1}{2} \s^c
\s_c h^{(0)}_{ab} + \s_a \s_b + \frac{3}{2} \s^2 h^{(0)}_{ab} = -\frac{2}{\sqrt{-h^{(0)}}}\frac{\delta L^{(\sigma)}}{\delta h^{(0)\:ab}}, \label{kab}
\eea
i.e., $\kappa_{ab}$ is precisely the stress-tensor of $L^{(\sigma)}$. See Appendices \ref{vecKilling} and \ref{classSTD}, and in particular around \eqref{kabapp}, for more properties of this tensor. Using integration by parts and properties of Killing vectors on the unit hyperboloid \eqref{eq:T1}-\eqref{eq:T2}, it follows that the integrability conditions \eqref{intO} are equivalent to
\be
\int_S d^2S \, \kappa_{ab}\xi_{(0)}^a n^b = 0.\label{BeigIntk}
\ee

\section{Mutually Dual Conserved Tensors}
\label{one}

In this section we present a general construction of conserved Lorentz charges. Our approach is to construct these charges using symmetric and divergence-free tensors. We construct these tensors from the equations of motion. Since Killing vectors corresponding to asymptotic Lorentz transformations are larger at infinity than translations, corresponding conserved tensors are constructed both from the leading and the next-to-leading terms in the Beig-Schmidt expansion.  These tensors are linear in the next-to-leading terms and quadratic in the leading terms.

Since Killing vectors on the hyperboloid obey the duality relations \eqref{dualxi}, any expression for conserved Lorentz charges involving derivatives of $\xi_{(0)}$  can be written in an equivalent form without derivatives acting on $\xi_{(0)}$ as
\bea
\int_S d^2 S T_{ab} \xi_{(0)}^{a} n^b
\eea
where $n^a$ is the unit normal to the sphere and $T_{ab}$ is symmetric. A sufficient condition for the charges to be conserved is that $T_{ab}$ should be divergence-free.  We have seen previously that one can form at least one symmetric, divergence-free, and traceless (SDT) tensor linear in $h_{ab}^{(2)}$, namely $E^{(2)}_{ab} - \s E^{(1)}_{ab}$ and another SDT tensor linear in the curl of  $h_{ab}^{(2)}$, namely $B^{(2)}_{ab}$. Now,  properties of Killing vectors on $dS_3$ detailed in Appendix \ref{vecKilling} imply the following equivalent representations of the conserved charges associated with boosts and rotations,
\bea
\mathcal J_{(i)} &\equiv & \frac{1}{8\pi G}\int_S d^2S (E^{(2)}_{ab} - \s E^{(1)}_{ab})\xi^a_{\rom{rot}(i)}n^b = - \frac{1}{8\pi G}\int_S d^2S  B^{(2)}_{ab} \xi^a_{\rom{boost}(i)}n^b ,\,\label{eq:J} \\
\mathcal K_{(i)} &\equiv & \frac{1}{8\pi G}\int_S d^2S (E^{(2)}_{ab} - \s E^{(1)}_{ab})\xi^a_{\rom{boost}(i)}n^b = \frac{1}{8\pi G} \int_S d^2S B^{(2)}_{ab} \xi^a_{\rom{rot}(i)}n^b  \label{eq:K}.
\eea
The conserved charges written in the second form using $B^{(2)}_{ab}$ are exactly the charges defined by Ashtekar-Hansen in \cite{Ashtekar:1978zz, AshRev}.

We now show that these conserved quantities are uniquely defined, in the sense that the addition to SDT tensors $E^{(2)}_{ab} - \s E^{(1)}_{ab}$ or  $B^{(2)}_{ab}$ of any symmetric and divergence-free tensor quadratic in $\sigma$ and its derivatives  does not modify the value of these conserved charges, once the integrability conditions are obeyed. Indeed, from the classification performed in Appendix \ref{classSTD}, any SDT tensor quadratic in $\sigma$ and its derivatives can be written as
\bea\label{XnSTD}
a_{(3)} (\curl \kappa)_{(ab)} + a_{(4)} (\curl^2 \kappa)_{ab} + a_{(5)} (\curl^3 \kappa)_{ab} +  a_{(6)} (\curl^4 \kappa)_{ab} + \dots,
\eea
for some coefficients $a_{(i)}$, with $\kappa_{ab}$ given in \eqref{kab}. The term $(\curl \kappa)_{(ab)}=-\eps_{cd(a}\s^c \s^d_{\;\, b)}$ does not contribute to the charges once the integrability conditions \eqref{intO} are imposed. Now, as proven in \eqref{eq:T1} and \eqref{eq:T2} in Appendix \ref{vecKilling}, charges associated to the symmetrized curl of a symmetric and divergence-free tensor can be related to the charges associated with the symmetric and divergence-free tensor itself. Using this argument iteratively, charges associated with each individual term in the expansion \eqref{XnSTD} are equivalent to the charges \eqref{intO}, which vanish as a consequence of the integrability conditions. The symmetric and divergence-free tensors with non-zero trace also do not contribute to charges. In this class, tensors for which the symmetrized curl is zero do not contribute  to the charges as a consequence of the relations \eqref{eq:T1} and \eqref{eq:T2}.  As shown in Appendix \ref{classSTD}, there is a unique independent symmetric and divergence-free tensor for which both the trace and the symmetrized curl are non-zero.  It is given by $\kappa_{ab}$ and does not contribute to the charges as shown in 
equation \eqref{BeigIntk}.

Since the six Lorentz charges are uniquely defined by \eqref{eq:J}-\eqref{eq:K}, all definitions proposed in the literature should reduce to these expressions. We show in the following section that the Ashtekar-Hansen charges are equivalent to the counter-term charges. We differ to a forthcoming publication \cite{Compere:2011bb} the comparison of these expressions to the standard ADM Hamiltonian conserved charges \cite{Dirac:1958jc,Deser:1960zzc,Arnowitt:1961zz, Arnowitt:1962aa,Regge:1974zd,Beig:1987aa}, to the charges defined from the linearized equations of motion \cite{Abbott:1981ff}, to the covariant phase space charges \cite{Lee:1990nz,Wald:1993nt,Iyer:1994ys} and to the cohomological methods \cite{Barnich:2001jy,Barnich:2003xg,Barnich:2007bf,Compere:2007az} and to charges with boundary counterterms contributions \cite{CompereMarolf}.

\section{Counter-term and Ashtekar-Hansen Charges Revisited}
\label{two}

In \cite{Mann:2006bd, Mann:2008ay} the equivalence between the counter-term charges and the Ashtekar-Hansen charges was established. Here we revisit this calculation and point out that the symmetry condition on $\sigma$ explicitly used in \cite{Mann:2008ay} is in fact not required. Integrability conditions mentioned above on $\sigma$ are precisely what is needed to show the equivalence between the Lorentz counter-term charges and the Ashtekar-Hansen charges. For translations there is no change as compared to the previous work, so here we are only concerned with Lorentz charges.

The boundary stress tensor (\ref{StressTensor}) obtained from the action (\ref{covaction}) can be used to define the conserved charge
 \begin{equation}\label{GenericCharge}
 Q[\xi] = \frac{1}{8 \pi G} \int_{S_\rho} d^2 S \sqrt{-h} \,  T_{ab} \,  u^a  \xi^{b},
\end{equation}
for any asymptotic Killing field $\xi^{a}$, where $h_{ab}$ is the induced metric on the hyperboloid $\mathcal{H}_\rho$ defined as a constant $\rho$ slice, $S_\rho$ is a Cauchy surface in $\mathcal H_\rho$ and $u^{a}$ is a timelike unit vector in $\mathcal{H}_\rho$ normal to $S_\rho$. Expanding the expression in powers of $\rho$ for rotations and boosts, one notices a potentially linearly divergent term in $\rho$. However, since $E_{ab}^{(1)}$ admits $\s$ as its scalar potential, 
the divergent term is in fact zero \cite{Ashtekar:1978zz}.
Then, the finite part of the stress-tensor \eqref{GenericCharge} reduces to \cite{Mann:2006bd, Mann:2008ay}
\bea
Q[\xi_{(0)}] = \frac{1}{8 \pi G} \int_{S} d^2 S \sqrt{-h_{(0)}}  (E_{ab}^{(2)}-\sigma E_{ab}^{(1)}- \gamma_{ab} -\Delta_{ab}^{(2)})\xi_{(0)}^a n^b \label{tough}
\eea
where $n^a$ is the leading order coefficient of $u^a$ in the asymptotic expansion and $S$ is the unit two-sphere.
It is now easy to show that $\gamma_{ab}$ does not contribute to conserved charges:  $n^{a} \gamma_{ab} \xi^b_{(0)}$ is a total divergence on $S$.  This is because,
\bea
\gamma_{ab} \xi_{(0)}^a &=&  D^a \left( \xi_{(0)}^c D_{[a} \kappa_{b]c} + \kappa_{c[a} D_{b]} \xi_{(0)}^c \right) +  D^a ( \s^2 D_{[a} \xi^{(0)}_{b]})  - 4 D^a ( \s \s_{[a} \xi^{(0)}_{b]}),
\eea
where $\kappa_{ab}$ is defined in \eqref{kab}.  
The tensor $\Delta^{(2)}_{ab}$ also does not contribute to conserved charges. This is because $\Delta^{(2)}_{ab}$ is an SDT tensor and can be written as in (\ref{XnSTD}).
Indeed,
\bea
\Delta^{(2)}_{ab} &=& -\frac{7}{4} (\curl^2 \kappa)_{ab}-\frac{3}{4} (\curl^4 \kappa)_{ab}.
\eea
Imposing integrability conditions, the result  immediately follows. Thus, expression \eqref{tough} simply reduces to
\bea
Q[\xi_{(0)}] = \frac{1}{8 \pi G} \int_{S} d^2 S \sqrt{-h_{(0)}}  (E_{ab}^{(2)}-\sigma E_{ab}^{(1)})\xi_{(0)}^a n^b,
\eea
which is identical to \eqref{eq:J}--\eqref{eq:K}.

We have thus demonstrated the explicit agreement between the Ashtekar-Hansen charges and the counter-term charges without assuming $\sigma$ to be symmetric, but requiring integrability of the Beig-Schmidt expansion.

\subsection*{Acknowledgements}
We would like to thank G. Barnich, S. Deser, I. Morrison, I. Papadimitriou, K. Skenderis, C. Troessaert and especially D. Marolf for useful discussions. GC acknowledges support from a `Nederlandse Organisatie voor Wetenschappelijk Onderzoek' (NWO) Vici grant. AV and FD were supported by IISN Belgium (conventions 4.4511.06 and 4.4514.08), and by the Belgian Federal Science Policy Office through the Interuniversity Attraction Pole P6/11.

\appendix
\section{Properties of Tensors on $dS_3$}

\label{proptensors}

In this appendix we first present the proof of two lemmae previously mentioned in the works of Ashtekar and Hansen \cite{Ashtekar:1978zz, AshRev} and of Beig and Schmidt \cite{BS} but not explicitly proven in these original references. We also state and prove three additional lemmae of general interest that are used as guidelines in the main text.

The metric on the unit three-dimensional de-Sitter space is
\be
ds_{\rom{dS}_3}^2 = h^{(0)}_{ab}dx^a dx^b = -d\tau^2 + \cosh^2 \tau (d\theta^2 + \sin^2 \theta d \phi^2),
\ee
where $-\infty < \tau < \infty$, $ 0 \le \theta \le \pi$ and $0 \le \phi \le 2 \pi$. The coordinates $(\theta, \phi)$ parameterize the unit two-sphere. The Riemann and Ricci tensors and Ricci scalar of the unit metric $h_{ab}^{(0)}$ on the hyperboloid are given by
\be
R^{(0)}_{abcd} = h^{(0)}_{ac} h^{(0)}_{bd} - h^{(0)}_{ad} h^{(0)}_{bc},\qquad R^{(0)}_{ab} = 2 h_{ab}^{(0)},\qquad R^{(0)} = 6.
\ee
It is useful to first note that in the harmonic decomposition on $dS_3$, the eight solutions of $\square \sigma +3\sigma  =0$ with harmonics $l=m=0$ or $l=1$, $m = -1,0,1$ are given by the four solutions
\bea
\zeta_{(0)} = \sinh\tau,\qquad \zeta_{(k)} = \cosh\tau f_{(k)}, \quad k=1,2,3,\label{zetas}
\eea
of $\DD_a \DD_b \zeta_{(a)} + h_{ab}^{(0)}\zeta_{(a)}= 0,$ which are odd under parity $(\tau, \theta, \phi) \rightarrow (-\tau, \pi-\theta, \phi+\pi)$ and the four functions
\begin{eqnarray}
\hat \zeta_{(0)} &=& \frac{\cosh 2\tau}{\cosh \tau}, \qquad \hat \zeta_{(k)} = \left(2\sinh \tau+\frac{\tanh \tau}{\cosh \tau}\right)f_{(k)}\label{zetahat} ,\quad k=1,2,3,
\end{eqnarray}
which are even under parity. Here,
\bea
f_{(1)} &=& \cos \theta, \qquad f_{(2)} =  \sin \theta \cos \phi, \qquad f_{(3)} =  \sin \theta \sin \phi , \label{harmosphere}
\eea
are the three $l=1$ harmonics on the two-sphere.

\begin{lemma}[Ashtekar-Hansen]\label{Ash} On the three-dimensional hyperboloid, any traceless curl-free divergence-free symmetric tensor $T_{ab}$ such that $\square T_{ab}= 3 T_{ab}$ can be written as
\begin{equation}
T_{ab} = \DD_a \DD_b \Phi + h^{(0)}_{ab}\Phi,
\end{equation}
with $\square \Phi + 3\Phi = 0$. The scalar $\Phi$ is determined up to the ambiguity of adding a combination of the four functions \eqref{zetas}. 
\end{lemma}

\begin{lemma}[Beig-Schmidt] \label{BS} On the three dimensional hyperboloid, any scalar $\Phi$ satisfying $\square \Phi + 3\Phi = 0$ and such that it does not contain the four lowest hyperbolic harmonics \eqref{zetahat} defines a symmetric, traceless, curl-free and divergence-free tensor $T_{ab} = \DD_a \DD_b \Phi + h^{(0)}_{ab}\Phi$ that can be written as
\begin{equation}
T_{ab} = \epsilon_{a}^{\;\, cd}\DD_c P_{d b},
\end{equation}
where $P_{ab}$ is a symmetric, traceless and divergence-free tensor. This tensor is defined up to the ambiguity $P_{ab} \rightarrow P_{ab} + \DD_a \DD_b \omega + h^{(0)}_{ab}\omega$ where $\omega$ is an arbitrary scalar obeying $\square \omega +3\omega =0$.
\end{lemma}

\begin{lemma}\label{lemma8}
On the hyperboloid, any regular symmetric divergence-free traceless tensor $T_{ab}$ obeying $(\square -2) \, T_{ab}=0$ can be uniquely decomposed as
\bea
T_{ab} = \sum_{i=1}^3 \left( v_{(i)} V_{(i)ab}+ w_{(i)} W_{(i)ab} \right)+ J_{ab},
\eea
where the three tensors $V_{(i)ab}$ and the three tensors $W_{(i)ab}$, $i=1,2,3$ are given by
\begin{align}
V_{(i)\tau\tau} &= 2 \sech ^5 \tau \zeta_{(i)},& V_{(i)\tau i} &= \sech ^3 \tau \tanh \tau \partial_i \zeta_{(i)},& V_{(i)ij}&=\eta_{ij} \sech ^3\tau\zeta_{(i)}, \\
W_{(i)\tau \tau} &= 0,&   W_{(i) \tau i} &=  \sech ^3\tau  \eps_{i}^{\;\, j}\partial_j \zeta_{(i)},&   W_{(i)ij} &=0.
\end{align}
These tensors are dual to each other in the sense
\bea
\eps_a^{\;\, cd}\DD_c V_{(i)d b} = -W_{(i)ab},\qquad \eps_a^{\;\, cd}\DD_c W_{(i)d b}=  V_{(i)ab}.
\eea
These tensors also obey the orthogonality properties
\begin{align}
\int_{S}  V_{(k)ab}\zeta^a_{(l)} n^b d^2S    =  \int_{S} W_{(k)ab}\zeta^a_{(l)} n^b d^2S  = 0,\\
\int_{S}  V_{(k)ab}\xi^a_{\rom{rot}(l)} n^b d^2S    =   \int_{S}  W_{(k)ab}\xi^a_{\rom{boost}(l)}n^b d^2S = 0,\\
\int_{S}  V_{(k)ab}\xi^a_{\rom{boost}(l)}n^b d^2S  = \int_{S}  W_{(k)ab}\xi^a_{\rom{rot}(l)} n^b d^2S    = \frac{8\pi}{3} \delta_{(k)(l)},
\end{align}
where $\delta_{(i)(j)} = 1$ if $i=j$, and $\zeta^a_{(l)} = D^a \zeta_{(l)}$ are the four translation Killing vectors (conformal Killing vectors on $dS_3$) with $\zeta_{(l)}$ given in \eqref{zetas}. The tensor $J_{ab}$ is a symmetric traceless divergence-free tensor obeying
\bea
\int_{S} d^2S J_{ab}\xi_\rom{rot}^a n^b  = \int_{S} d^2S J_{ab} \xi_\rom{boost}^a n^b =\int_{S} d^2S J_{ab}\zeta^a_{(l)} n^b  = 0, \qquad (\square -2)J_{ab}= 0.
\eea
\end{lemma}

\begin{lemma} \label{newlemma} On the three dimensional hyperboloid, any symmetric traceless and divergence-free tensor can be decomposed as
\bea
T_{ab} = \curl(\tilde T_{ab}) + \DD_a \DD_b \hat \zeta+h_{ab}^{(0)}\hat \zeta
\eea
where $\tilde T_{ab}$ is a symmetric, traceless and divergence-free tensor and $\hat\zeta$ is a combination of the four functions \eqref{zetahat}.
\end{lemma}

\begin{lemma} \label{higheroplemma} On the hyperboloid, any regular symmetric divergence-free traceless tensor $T_{ab}$ obeying $(\square +n^2 -2n-2 ) \, T_{ab}=0$ with $n$ any integer $n \geq 3$ also obeys
\bea
\int_{S}  T_{ab}\zeta^a_{(l)} n^b d^2S  = 0, \; \; l=0,1,2,3   , \qquad \int_{S}  T_{ab}\xi^a_{(0)} n^b d^2S = 0  \, ,
\eea
where  $\zeta^a_{(l)} = D^a \zeta_{(l)}$ are the four translation Killing vectors (conformal Killing vectors on $dS_3$) with $\zeta_{(l)}$ given in \eqref{zetas} and $\xi^a_{(0)}$ are the six Killing vectors on $dS_3$. 
\end{lemma}


Before starting the proof of these lemmae, we recall the following properties of tensors on $dS_3$.
Let $t$, $t_a$, $t_{ab} = t_{(ab)}$ be some arbitrary fields on the hyperboloid, then
\bea
\left[ \DD_a,\square \right] t &=& -2 \DD_a t,\\
\left[ \DD_a,\DD_b \right]t_{c} &=& h^{(0)}_{ac}t_b - h^{(0)}_{bc}t_a,\\
\left[ \DD_a,\square \right]t_b &=& 2 h^{(0)}_{ab} \DD_c t^c - 4 \DD_{(a} t_{b)},\\
\left[ \DD^c,\DD_a \right] t_{cb} &=& 3 t_{ba} - h_{ab}^{(0)} t^c_{\; c},\\
\left[ \DD_a,\DD_b \right] t_{cd} &=& 2 h^{(0)}_{a(c} t_{d)b} - 2 h^{(0)}_{b(c} t_{d)a},\\
\left[ \DD_a,\square \right] t_{bc} &=& 4 h^{(0)}_{a(b}\DD^d t_{c)d} -6\DD_{(a}t_{bc)},\\
\left[ \DD^b,\square \right] t_{bc} &=& 4 \DD^d t_{cd} -2 \DD_c t .
\eea
These identities are useful in several arguments in the main text and in the proof of the lemmae.

\vskip 0.2cm

\noindent {\bf Proofs of lemmae \ref{Ash}, \ref{BS}, \ref{lemma8}, \ref{newlemma} and \ref{higheroplemma}:} The five lemmae have an overlapping proof.  In order to establish these lemmae we need to derive decomposition of regular symmetric, divergence-free, and traceless (SDT) tensors on the hyperboloid. We do so in the rest of this appendix.

A general regular symmetric tensor on the hyperboloid can be expressed as a linear combination of symmetric tensors built from two-dimensional spherical harmonics. A general such tensor has the form
\bea
T_{\tau \tau} &=& f_1(\tau) Y_{lm}(\theta,\phi),\\
T_{\tau i} &=& f_2 (\tau)D^{(2)}_i Y_{lm}(\theta,\phi) + f_3(\tau) \eps_i^{\;\, j}D^{(2)}_j Y_{lm}(\theta,\phi),\\
T_{ij} &=& f_4(\tau)\left(D^{(2)}_i D^{(2)}_j +\frac{l(l+1)}{2} \eta_{ij} \right) Y_{lm}(\theta,\phi)   + f_5(\tau) \eps_{(i}^{\;\, k}D^{(2)}_{j)}D^{(2)}_k Y_{lm}(\theta,\phi) \nn  \\ & & + f_6(\tau) \eta_{ij} Y_{lm}(\theta,\phi),
\eea
where indices $i,j,k$ run over two-sphere $(\theta, \phi)$, $Y_{lm}(\theta,\phi)$ are scalar spherical harmonics on the two sphere with $l=0, 1, \ldots,$ and $m=-l, \ldots, l$, and  $D^{(2)}_i$ is the covariant derivative compatible with the round metric $\eta_{ij}$ on the two-sphere. In writing these expression we have already made use of the identity
\be
\left( D^{(2)}_i D^{(2)}_j  + \frac{l(l+1)}{2}\eta_{ij} + \eps_{(i}^{\;\, k}\eps_{j)}^{\;\, l}D^{(2)}_k D^{(2)}_l \right) Y_{lm}(\theta,\phi) = 0,
\ee
in order to reabsorb the tensor structure $\eps_{(i}^{\;\, k}\eps_{j)}^{\;\, l}D^{(2)}_k D^{(2)}_l  Y_{lm}$ into the definition of $f_4(\tau)$. The tensor is traceless if and only if
\be
f_1(\tau) = 2 \sech^2\tau f_6(\tau).
\ee
For the case $l=0$, $m=0$, only $f_6(\tau)$ parameterizes non-zero tensors. The divergence-free condition is solved only for $f_6 \sim \sech\tau$. The general tensor then reduces to
\be
T_{ab} = \DD_a \DD_b \hat \zeta_{(0)} + h_{ab}^{(0)}\hat  \zeta_{(0)}.
\ee
When $l=1$, we have that $ \eps_{(i}^{\;\, k}D^{(2)}_{j)}D^{(2)}_k Y_{lm}(\theta,\phi) = 0$. Therefore, $f_4$ and $f_5$ do not lead to non-zero tensors. The divergence-free condition fixes
\bea
f_2(\tau) &=& -\tanh\tau f_6(\tau) - \partial_\tau f_6(\tau), \\
f_3(\tau)&=& C_1 \sech^2\tau,\\
f_6(\tau)&=& C_2 \sech^2\tau +C_3 \sech\tau \tanh\tau,
\eea
where $C_1$, $C_2$ and $C_3$ are constants. There are therefore three solutions for each value of $m = -1,0,1$, so $9$ solutions in total. Three independent solutions are the tensors admitting a scalar potentials $\hat\zeta_{(i)}$, $i=1,2,3$,
\be
T_{ab} = \DD_a \DD_b\hat  \zeta_{(i)} + h_{ab}^{(0)} \hat \zeta_{(i)}.\label{Tzetai}
\ee
These tensors are curl-free and $(\square - 3)T_{ab} = 0$. The six other tensors can be written as a linear combination of the following two sets of three tensors,
\bea
V_{(k)\tau\tau} &=& 2\sech^5\tau \zeta_{(i)},\quad V_{(k)\tau i} = \sech^3\tau \tanh \tau D^{(2)}_i \zeta_{(k)},\quad V_{(k)ij}=\eta_{ij}\sech^3\tau\zeta_{(k)}, \\
W_{(k)\tau \tau} &=& 0,\quad W_{(k) \tau i} = \sech^3\tau  \eps_{i}^{\;\, j}D^{(2)}_j \zeta_{(k)},\quad W_{(k)ij} =0,
\eea
where $\zeta_{(k)} = \cosh\tau f_{(k)}$, $k = 1,2,3$. These tensors are dual to each other in the sense
\bea
\eps_a^{\;\, cd}D_c V_{d b} = -W_{ab},\qquad \eps_a^{\;\, cd}D_c W_{d b}=  V_{ab}.
\eea
Since applying two times the curl operator on a traceless, divergence-free, symmetric tensor is equivalent to applying $(\square -3)$, we deduce that both tensors obey
\bea
(\square -2)V_{ab} = 0,\qquad (\square -2)W_{ab} = 0.
\eea
These tensors also obey the orthogonality properties
\bea
\int_{S}V_{(k)ab}\xi^a_{\rom{rot}(l)} n^b d^2 S= 0, & \qquad & \int_{S} W_{(k)ab}\xi^a_{\rom{boost}(l)}n^b d^2 S= 0,\\
\int_{S} V_{(k)ab}\xi^a_{\rom{boost}(l)} n^b d^2 S = \frac{8\pi}{3} \delta_{(k)(l)}, &&
\int_{S} W_{(k)ab}\xi^a_{\rom{rot}(l)} n^b d^2 S = \frac{8\pi}{3} \delta_{(k)(l)},
\eea
where $C$ is a cut of the hyperboloid. Since $V_{(k)ab}$ and $W_{(k)ab}$ are divergence-free, these integrals are independent of the chosen cut of the hyperboloid.

For $l > 1$, we can solve the divergence-free condition in terms of $f_2$, $f_4$ and $f_5$ as
\bea
f_2(\tau) &=& -\frac{2}{l(l+1)}(\tanh \tau f_6(\tau) + \partial_\tau f_6(\tau)),\\
f_4(\tau) &=& \frac{2}{(l-1)l(l+1)(l+2)}\Big( (l+l^2+2\cosh 2\tau)f_6(\tau) \nn \\
&& +2\cosh\tau (3\sinh\tau \partial_\tau f_6(\tau)+\cosh\tau \partial_\tau \partial_\tau f_6(\tau) )\Big),\\
f_5(\tau)&=& - \frac{2}{(l-1)(l+2)}\cosh\tau (2\sinh\tau f_3(\tau)+\cosh\tau \partial_\tau f_3(\tau)).
\eea
The general tensor with harmonics $(l,m)$, $l>1$ is a linear combination of the following two tensors depending each on an arbitrary function $f(\tau)$ of $\tau$,
\bea
T_{ab}^{(I)}(f) & \equiv & T_{ab}\left(f_3(\tau) = \frac{1}{l(l+1)} f(\tau),\,f_6(\tau) = 0\right),\label{gen45}\\
T_{ab}^{(II)}(f) & \equiv & T_{ab}\left(f_3(\tau) =0,\,f_6(\tau) = \frac{1}{2} f(\tau) \right).\label{gen46}
\eea
These tensors obey the remarkable properties
\bea
\eps_a^{\;\, cd}\DD_c T_{db}^{(I)}(f) &=& T_{ab}^{(II)}(f),\label{curlTI}\\
\eps_a^{\;\, cd}\DD_c T_{db}^{(II)}(f) &=& - T_{ab}^{(I)}( \mathcal O f),\label{curlTI2}
\eea
where $ \mathcal O f$ is the following differential operator acting on $f(\tau)$,
\be
\mathcal O f \equiv (1+l(l+1)\sech^2\tau)f+2\tanh\tau \partial_\tau f+\partial_\tau \partial_\tau f\, .
\ee
We deduce also the following properties
\bea
(\square - 3)  T_{ab}^{(I)}(f) &=& - T_{ab}^{(I)}(\mathcal O f), \label{eq:box1}\\
(\square - 3)  T_{ab}^{(II)}(f) &=& - T_{ab}^{(II)}(\mathcal O f).\label{eq:box2}
\eea
Using the explicit expression for the tensors and the orthogonality of spherical harmonics, we also have
\bea
\int_ST_{ab}^{(I)}(f) \xi_{\rom{rot}(k)}^a n^b d^2 S =0, &\qquad  &\int_S T_{ab}^{(II)}(f) \xi_{\rom{rot}(k)}^a n^b d^2 S =0, \label{killJ1} \\
\int_S T_{ab}^{(I)}(f) \xi_{\rom{boost}(k)}^a n^b d^2 S =0,  &\qquad  &  \int_S T_{ab}^{(II)}(f) \xi_{\rom{boost}(k)}^a n^b d^2 S =0. \label{killJ2}\\
\int_S T_{ab}^{(I)}(f) \DD^a \zeta_{(l)} n^b d^2 S =0,  &\qquad  &  \int_S T_{ab}^{(II)}(f) \zeta_{(l)}^a n^b d^2 S =0. \label{killJ3}
\eea
The above decomposition proves lemma \ref{lemma8}. Indeed, one can isolate the $l=0,1$ harmonics and then all the higher harmonics can be regrouped in a tensor $J_{ab}$ that obeys $\int_S J_{ab} \xi_\rom{rot}^a n^b=\int_S J_{ab} \xi_\rom{boost}^a n^b=0$ as a consequence of \eqref{killJ1}-\eqref{killJ2}.

There are two special sets of two functions $f(\tau)$: the ones for which the differential operator obeys $\mathcal O f_{(0)}=0$ and the others for which $\mathcal O f_{(1)}= f_{(1)}$. The two functions obeying $\mathcal O f_{(0)}=0$ define tensors $T_{ab}^{(II)}(f_{(0)})$ such that
\bea
\DD_{[a}T_{b]c}^{(II)}(f_{(0)}) =0,\qquad (\square - 3) T_{ab}^{(II)}(f_{(0)})  = 0.
\eea
The tensor $T_{ab}^{(I)}(f_{(0)})$ is a tensor potential for $T_{ab}^{(II)}(f_{(0)})$ and is uniquely determined for the two solutions of $\mathcal O f_{(0)} =0$. From the explicit form of the tensor, we note that $T_{ab}^{(II)}(f_{(0)})$ can be written as
\bea
T_{ab}^{(II)}(f_{(0)}) = \DD_a \DD_b \Phi + h_{ab}^{(0)}\Phi, \label{scd2}
\eea
where $\Phi = \sum_{l}\sum_{m= -l}^{l}\Phi^{lm}(\tau) Y_{lm}(\theta,\phi)$ is a scalar that obeys $(\square + 3)\Phi =0$. The two independent solutions of $\mathcal O f_{(0)}=0$ correspond to the two independent solutions of the equation $(\square + 3)\Phi =0$ for fixed values of $l>1$, $-l \leq m \leq l$.

The two independent solutions for $f(\tau)$ of the differential equation $\mathcal O f_{(1)}=f_{(1)}$ can be used to define two pairs of dual tensors,
\bea
W_{ab} = T_{ab}^{(I)}(f_{(1)}) , \qquad V_{ab} = T_{ab}^{(II)}(f_{(1)}),
\eea
which obey
\bea
\eps_a^{\;\, cd}\DD_c V_{db}  = -W_{ab}, \qquad  \eps_a^{\;\, cd}\DD_c W_{db}  = V_{ab}\\
(\square -2)V_{ab} = 0,\qquad (\square -2)W_{ab} = 0.
\eea
Given the special role of the eigenfunction of  the operator $\mathcal O$, it is natural to decompose the functions $f(\tau)$ in that basis. The equation
\be
\mathcal O f_{(n)} = (n-1)^2 f_{(n)}\label{eqgen7}
\ee
for each positive integer $n$ is solved by associated Legendre functions of the first and second kind,
\be
f^{(1)}_{(n)} = \sech\tau P_l^n[\tanh \tau],\qquad  f^{(2)}_{(n)} = \sech\tau Q_l^n[\tanh \tau].
\ee
Lemma \ref{Ash} is then proven as follows. A symmetric traceless divergence-free tensor obeying $(\square - 3)T_{ab}=0$ can be decomposed into harmonics. The only possible tensors in harmonics $l =0,1$ have the form
\be
T_{ab} = \DD_a \DD_b \Phi + h_{ab}^{(0)} \Phi, \label{scd}
\ee
where $(\Box +3)\Phi = 0$ contains $l=0,1$ harmonics. For $l > 1$, we have seen that any tensor can be decomposed as a combination of two different tensor structures $T^{(I)}_{ab}(f^{(I)})$ and $T^{(II)}_{ab}(f^{(II)})$ depending each on one function. We then see from  \eqref{eq:box1}-\eqref{eq:box2} that such tensors obey $(\square - 3)T_{ab} =0$ if and only if $f^{(I)} = f^{(II)} = f_{(0)}$ where $f_{(0)}$ are the solutions of the differential equation $\mathcal O f_{(0)} =0$. Then, we note using \eqref{curlTI} that $T^{(I)}_{ab}(f_{(0)})$ is not curl-free and thus does not obey the preconditions of the lemma. The only remaining tensors have the form $T^{(II)}_{ab}(f_{(0)})$ and they can be written in terms of a scalar potential \eqref{scd2} as shown earlier.

The general solution of $(\Box +3)\Phi = 0$ contains $\zeta_{(i)}$, $\hat \zeta_{(i)}$, $i=0,1,2,3$ and the higher $l>1$ harmonics. For each value of $l>1$, $m$, there are two solutions for $\Phi$ that uniquely correspond to the two tensors $T^{(II)}_{ab}(f_{(0)})$. The four lower harmonics correspond to the tensor $T_{ab}$ built in \eqref{Tzetai}. The dependence in $\zeta_{(i)}$ is arbitrary since these scalars can then be added to $\Phi$ without changing $T_{ab}$. This ends the proof of lemma \ref{Ash}.

The lemma \ref{BS} is proven by noticing that by lemma \ref{Ash}, all tensors derived from a scalar using \eqref{scd} that have $l>1$ harmonics have the form $T^{(II)}_{ab}(f_{(0)})$. The tensor $T^{(I)}_{ab}(f_{(0)})$ is then the tensor potential for $T^{(II)}_{ab}(f_{(0)})$ by  \eqref{curlTI}.

Let us now prove lemma \ref{newlemma}. We consider an arbitrary SDT tensor $T_{ab}$. One can decompose it in $l=0$, $l=1$ and $l>1$ harmonics, and further the arbitrary functions $f(\tau)$ appearing in \eqref{gen45}-\eqref{gen46} can be decomposed in eigenfunctions \eqref{eqgen7} with positive integer $n$. The $l=0$, $l=1$ and $l>1$ harmonics with $n = 1$ can be written as the sum of a tensor admitting a scalar potential and the curl of an SDT tensor. From \eqref{curlTI}-\eqref{curlTI2}, the $l>1$ harmonics with $n > 1$ are explicitly the curl of an SDT tensor. Using in addition Lemma \ref{BS}, we obtain that $T_{ab}$ can be written as a sum of the curl of an SDT tensor and a sum of $\DD_a \DD_b \hat \zeta_{(i)} + h_{ab}^{(0)}\hat\zeta_{(i)}$, which proves the lemma.

Let us finally prove lemma \ref{higheroplemma}. Note that no SDT tensor obeying $(\square +n^2 - 2n -2)T_{ab} = 0$ with $n > 2$
integer can contain spherical harmonics $l =0$ or $l = 1$. This follows from the explicit form of the $l=0$ and $l=1$ SDT harmonics presented above. Therefore, any SDT tensor obeying $(\square +n^2 - 2n -2)T_{ab} = 0$ with $n > 2$ can be decomposed in the basis of tensors $T_{ab}^{(I)}(f)$ and $T_{ab}^{(II)}(f)$ \eqref{gen45}-\eqref{gen46} for $f(\tau)$ obeying the eigenvalue equation \eqref{eqgen7}. All such tensors are expanded in spherical harmonics with $l>1$.  The lemma then follows from the orthogonality of spherical harmonics \eqref{killJ1}--\eqref{killJ3}.

\section{Properties of Killing Vectors on $dS_3$}
\label{vecKilling}

Three-dimensional de-Sitter space admits six Killing vectors. Three of them are rotations and the other three correspond to four-dimensional Lorentz boosts when interpreted in the asymptotically flat context. The rotations are
\bea
\xi^{a}_{\rom{rot}(1)} \partial_a &=& \partial_\phi, \\
\xi^{a}_{\rom{rot}(2)} \partial_a &=& - \sin \phi \partial_\theta - \cot \theta \cos \phi \partial_\phi,  \\
\xi^{a}_{\rom{rot}(3)} \partial_a &=& \cos \phi \partial_\theta - \cot \theta \sin \phi \partial_\phi\, .
\eea
These Killing vectors are precisely the three Killing vectors of the round two-sphere. On the round two-sphere, Killing vectors satisfy a special property that they can be written as
\be \xi^{m}_{\rom{rot}(k)} = \epsilon_{(2)}{}^{mn} D_{n}^{(2)} f_{(k)}, \label{rotexpl}
\ee
where $f_{(k)}$ are the three scalar $l = 1$ harmonics on the two-sphere
\bea
(\square^{(2)} + 2)f_{(k)} = 0,\qquad \int_S d^2 S f_{(k)} f_{(l)} = \frac{4\pi}{3}\delta_{(k)(l)},
\eea
given explicitly in \eqref{harmosphere}. We use the conventions $\epsilon_{(S^2)}{}_{\theta \phi} = \sin \theta$, $D_{a}^{(2)}$ is the unique torsion free covariant derivative on $S^2$,  $dS = \sin\theta d\theta \wedge d\phi$, and $\square^{(2)}$ is the scalar Laplacian on $S^2$.

The boost Killing vectors of the three-dimensional de-Sitter space can be written as
\bea
\xi^{a}_{\rom{boost}(i)} = f_{(i)} n^{a} + \cosh\tau \sinh\tau h_{(0)}^{ab}\partial_b f_{(i)} \label{boostexpl}
\eea
or, explicitly, as
\bea
\xi^{a}_{\rom{boost}(1)} \partial_a &=& \cos \theta \partial_\tau - \tanh \tau \sin \theta  \partial_\theta, \\
\xi^{a}_{\rom{boost}(2)} \partial_a &=& \sin \theta \cos \phi \partial_\tau +  \tanh \tau \cos \theta \cos \phi  \partial_\theta - \tanh \tau  \csc \theta \sin \phi \partial_\phi,\\
\xi^{a}_{\rom{boost}(3)} \partial_a &=&  \sin \theta \sin \phi \partial_\tau +  \tanh \tau \cos \theta \sin \phi  \partial_\theta + \tanh \tau  \csc \theta \cos \phi \partial_\phi.
\eea
The unit vector normal to the two sphere in $dS_3$ is $n^{a}\partial_a = \partial_\tau$.

The boost Killing vectors are intimately related to the rotational Killing vectors by the following relation
\bea
\xi^{a}_{\rom{boost}(i)} &=& -\frac{1}{2}\eps^{abc} \DD_b \xi_{\rom{rot}(i)c},\\
\xi^{a}_{\rom{rot}(i)} &=& \frac{1}{2}\eps^{abc} \DD_b \xi_{\rom{boost}(i)c},
\eea
where $\DD_a$ is the covariant derivative on the hyperboloid and $\eps_{abc}$ the totally anti-symmetric tensor normalized as $\eps_{\tau \theta \phi} = +\cosh^2\tau \sin\theta$. The latter relation implies
\bea
(\square +2)\xi^{a}_{\rom{rot}(i)} =0,\qquad (\square +2)\xi^{a}_{\rom{boost}(i)} =0,
\eea
where $\square = \DD^a\DD_a$.

Finally let us derive how these relations imply two equivalent forms for the conserved charges associated with boosts and rotations. Given a tensor  $T_{ab}$ without special properties, one can show that on the $\tau = 0$ slice of de Sitter space,
\bea
T_{ab}\xi_{\rom{rot} (i)}^a n^b &=& \eps_a^{\; cd}\DD_c T_{db}\xi^a_{\rom{boost}(i)}n^b+\DD^{(S^2)}_c ( T_{ab}\eps_{(S^2)}^{ac} f_{(i)} n^b)\, .\label{proprotboost}
\eea
For any symmetric and divergence-free tensor, one has $$\DD^b (T_{ab}\xi_{\rom{rot}(i)}^a) = 0, \qquad \qquad \DD^b ((\curl T)_{(ab)}\xi_{\rom{rot}(i)}^a) = 0.$$ Note that $(\curl T)_{ab}$ is symmetrized in the second equation, as $(\curl T)_{ab}$ is not necessarily symmetric. Therefore, for any regular symmetric and divergence-free tensor, the conserved charges
\bea
Q[T_{ab},\xi_{\rom{rot}(i)}^a]\equiv \int_S d^2 S \:  n^b T_{ab}\xi_{\rom{rot}(i)}^a.
\eea
can be expressed in two equivalent ways as follows,
\bea
Q[T_{ab},\xi_{\rom{rot}(i)}^a]= \int_S d^2 S \:  n^b T_{ab}\xi_{\rom{rot}(i)}^a = \int_S d^2 S \:  n^b \curl(T)_{(ab)} \xi_{\rom{boost}(i)}^a \, . \label{eq:T1}
\eea

Replacing $T_{ab}$ by the curl of $T_{ab}$ in identity \eqref{proprotboost}, we get on the $\tau = 0$ slice
\be
T_{ab}\xi_{\rom{boost}(i)}^b n^a =- \curl( T)_{ab}\xi^b_{\rom{rot}(i)}n^a+\DD^{(S^2)}_c (2\DD_{[d}T_{c]b} n^d n^b f_{(i)})+(\curl(\curl T) +T)_{ab}n^a \xi_{\rom{boost} (i)}^b. \nonumber
\ee
For any  symmetric and divergence-free tensor, we have $(\curl(\curl T) +T )_{ab}= (\square - 2)T_{ab} + h_{ab}^{(0)}T$, and
\bea
\left( (\square - 2)T_{ab} + 2 h_{ab}^{(0)} T \right) \xi_{(0)}^c = 2 \DD^a \left( \xi_{(0)}^c \DD_{[a} T_{b]c}+ T_{c[a}\DD_{b]}\xi_{(0)}^c \right).
\eea
Therefore, one obtains
\bea
\int_S d^2 S \: n^b T_{ab}\xi_{\rom{boost}(i)}^a =- \int_S d^2 S \:  n^b \curl(T)_{ab} \xi_{\rom{rot}(i)}^a - \int_S d^2 S \,  T \xi^{\rom{boost}(i)}_a n^a  \, .
\eea
For a tensor $T_{ab}$ whose trace is non-vanishing, $\curl(T)_{ab}$ is not symmetric in general. Decomposing into symmetric and anti-symmetric parts and using integrations by parts, the following conserved charges
\bea
Q[T_{ab},\xi_{\rom{boost}(i)}^a]\equiv \int_S d^2 S \:  n^b T_{ab}\xi_{
\rom{boost}(i)}^a. 
\eea
can also be expressed in two equivalent forms
\bea
Q[T_{ab},\xi_{\rom{boost}(i)}^a]= \int_S d^2 S \:  n^b T_{ab}\xi_{
\rom{boost}(i)}^a =- \int_S d^2 S \:  n^b \curl(T)_{(ab)} \xi_{\rom{rot}(i)}^a  \, .\label{eq:T2}
\eea
In establishing this, we used $\curl(T)_{[ab]}= -\frac{1}{2} \epsilon_{abc} \DD^c T$.

To summarize, we have shown in equations (\ref{eq:T1}) and (\ref{eq:T2}) that charges associated with any symmetric and divergence-free tensor are equivalent to charges associated with the symmetrized curl of this tensor. In particular, this means that charges constructed using an SDT tensor are equivalent to charges constructed using the curl of this SDT tensor. Also, charges constructed with symmetric and divergence free tensors that have zero symmetrized curl are automatically zero.

To finish this Appendix, let us briefly look at charges associated with translations.
Given a symmetric tensor $U_{ab}$, it is easy to show that
\bea
\curl(U)_{ab} \zeta_{(i)}^a n^b = \DD_c \left( \eps_b{}^{\; cd}U_{ad} \zeta_{(i)}^a \right) n^b. \label{curlzeta}
\eea
The r.h.s of this equation is a total divergence on the two-sphere. From Lemma \ref{newlemma} and \eqref{curlzeta} it then follows that  charges associated with translations constructed using  an SDT tensor $T_{ab}$ are simply associated with the coefficients of the four lowest harmonics $\hat \zeta_{(i)}$ given in \eqref{zetahat},
\bea
Q[T_{ab},\zeta_{(i)}^a]\equiv\int d^2S T_{ab}\zeta_{(i)}^a n^b = \int d^2S (\DD_a \DD_b \hat \zeta + h_{ab}^{(0)}\hat \zeta ) \zeta_{(i)}^a n^b\, .
\eea

\section{Classification of Symmetric and Divergence-free Tensors}
\label{classSTD}

In this appendix we classify all Symmetric and Divergence-free (SD) tensors that one can build from a quadratic expression in $\s$ and its derivatives. We  denote tensors $X^{[n]}_{ab}$ with a superscript $[n]$ indicating the highest number of derivatives of $\s$ that the tensor contains. We first prove by induction in $n$ that there is one and only one Symmetric, Divergence-free, and Traceless (SDT) tensor at each order $n \geq 3$ and none for $n \leq 2$.  This fact is then used to prove that at each odd order $n \geq 3$ any SD tensor is also SDT and that at each even order $n \geq 4$ any SD tensor is a linear combination of the unique SDT tensor and a tensor with vanishing curl. At order $n = 2$ the general SD tensor is a linear combination of a tensor whose symmetrized curl is SDT and a tensor whose symmetrized curl is zero.

Let us begin with some definitions. A large class of SDT tensors can be formed by taking the curl of symmetric tensors  $M_{ab}$ obeying $\DD^b M_{ab} = \DD_a M$. Such SDT tensors will be called tensors admitting a tensor potential. There can be SDT tensors that do not admit tensor potentials. Such tensors will be called irreducible SDT tensors. An SDT tensor is therefore a combination of an irreducible SDT tensor and an SDT tensor admitting a tensor potential. Non-trivial tensor potentials are the ones for which the curl is non-vanishing. Primitive tensor potentials are tensor potentials that are not curl of another tensor potential. There is an equivalence class of tensor potentials that defines the same SDT tensor. Indeed, tensors potentials with vanishing curl can be added to a non-trivial tensor potential but still define the same SDT tensor. We are only interested in Representatives of Non-trivial Tensor potentials in these equivalence classes, or RNT potentials in short. In summary, an SDT tensor is a combination of irreducible SDT tensors and curl of RNT potentials.

Let us now start the inductive proof. A generic symmetric tensor containing zero, one or two derivatives has the form
\bea
(a \sigma^2 +b \s_c \s^c )h_{ab}^{(0)} + c\s_a \s_b + d\s \s_{ab},
\eea
for some coefficients $a,b,c,d$. One can easily show that there are no SDT tensors in this class. However, there do exist tensor potentials of the form
\bea
P^{[2]}_{ab} \equiv \hat a \left(  (5 \s^2 +\s_d \s^d ) h_{ab}^{(0)} + 4\s \s_{ab} \right) + \hat b ( \DD_a \DD_b + h_{ab}^{(0)})\sigma^2 . \label{Mpreprimitive1}
\eea
The curl of the second term in this expression is zero. Therefore, the unique RNT potential (up to normalization) with at most two derivatives is given by \eqref{Mpreprimitive1} with  $\hat a = 1$, $\hat b = 0$,
\bea
M^{[2]}_{ab} =(5 \s^2 +\s_d \s^d ) h_{ab}^{(0)} + 4\s \s_{ab} . \label{Mprimitive1}
\eea
Since this potential cannot be obtained as the curl of another RNT potential, it is a primitive RNT potential. From this primitive RNT potential, one can build an SDT tensor with three derivatives by applying the symmetrized curl,
\bea
X^{[3]}_{ab} \equiv \eps_{(a}^{\;\,\; cd}\DD_{|c} M^{[2]}_{|d|b)} = 4 \eps_{cd(a}\s^c \s^d_{\;\, b)} = -4 \eps_{cd(a}\s^c E^{(1)}{}^d{}_{b)}\, .
\eea
Note that $X^{[3]}_{ab}$ is the right hand side of equation \eqref{B23eq} and it also appears in the integrability conditions \eqref{intO}. By a recursive application of the curl operator, one can build one SDT tensor at each order in derivatives. At next order, we have
\beqs
X^{[4]}_{ab}&=& \curl X^{[3]}_{ab}= (\square-3)M^{[2]}_{ab}-\DD_a \DD_b M^{[2]} + M^{[2]} h^{(0)}_{ab} \nonumber \\
&=& 2\sigma_c\sigma^c h^{(0)}_{ab} +2\sigma_{cd} \sigma^{cd} h^{(0)}_{ab} +2\sigma_{abc} \sigma^c -18 \sigma^2 h^{(0)}_{ab}-18 \sigma \sigma_{ab} -6 \sigma^c_{\;(a} \sigma_{b)}^{\:\;c}.
\eeqs
Note that $X^{[4]}_{ab}$ is the right hand side of equation \eqref{E23eq}.

Now, it is easily proven that the only tensor structure $X^{[2n+1]}$ is
\bea
\eps^{e f }{}_{(a} \s_{b) e c_1 c_2 \dots c_{n-1}} \s_f^{\; c_1 c_2 \dots c_{n-1}}\label{odd:dd}
\eea
up to terms containing $\s$ with less derivatives than $2n+1$. Using this tensor structure an SDT tensor can be constructed. Therefore, there can be at most one independent SDT tensor at order $[2n+1]$, which is the one we would find by applying successive curls on $M^{[2]}_{ab}$.

An SDT tensor with an even number of derivatives $[2n]$ has the form
\bea
 a h_{ab}^{(0)}\s_{c_1c_2 \dots c_n}\s^{c_1 c_2 \dots c_n}+b \s_{ab c_1 \dots c_{n-1}}\s^{c_1 \dots c_{n-1}}+c \s_{a c_1 \dots c_{n-1}}\s_b^{\; c_1\dots c_{n-1}}\label{even:dd}
\eea
plus lower derivative terms. For $n >1$, tracelessness and divergence free conditions require, respectively,
\bea
3 a +c = 0, \qquad  2a+b+c = 0.
\eea
As a result there is at most one SDT tensor $X^{[2n]}$ for $n>1$. Therefore, we conclude that there is one and only one SDT tensor for each $n \geq 3$ and none for $n \leq 2$. These are precisely the ones constructed by applying successive curls on the primitive RNT potential \eqref{Mprimitive1}. In particular, there are no irreducible SDT tensors built from quadratic terms in $\s$. Therefore, the general SDT tensor has the form
\bea\label{XnSTDapp}
 a_{(3)} X^{[3]}_{ab} + a_{(4)} \curl(X^{[3]})_{ab} + a_{(5)} \curl^2(X^{[3]})_{ab} +  a_{(6)}  \curl^3(X^{[3]})_{ab} + \dots
\eea
for some coefficients $a_{(i)}$.

Finally, let us classify symmetric and divergence-free (SD) tensors with non-zero trace. With at most two derivatives, one can construct from  \eqref{Mpreprimitive1} the SD tensor $P_{ab}^{[2]} - h_{ab}^{(0)} P_{cd}^{[2]}h_{(0)}^{cd}$. The symmetrized curl of this tensor is zero when $\hat a = 0$. Therefore, any SD tensor, with at most two derivatives, is the linear combination of a tensor whose symmetrized curl is zero and of the SD tensor
\bea\label{kaabb}
\kappa_{ab} = \s_a \s_b + h_{ab}^{(0)} \left(-\frac{1}{2}\s_c \s^c + \frac{3}{2}\s^2\right)\,\label{kabapp}
\eea
where we chose $\hat a = -1/4$ and $\hat b = 1/2$ in \eqref{Mpreprimitive1} for convenience. At orders $[2n+1]$, $n>1$, there is a unique tensor structure \eqref{odd:dd} leading to an SDT tensor $X^{[2n+1]}$. In particular, there is therefore no further independent SD tensor. At even orders $[2n]$, with $n>1$, one can form two independent SD tensors, \eqref{even:dd} with $2a+b+c=0$. One can write such a tensor as the SDT tensor $X^{[2n]}$ plus a tensor whose symmetrized curl is zero. Indeed, if the symmetrized curl of the independent tensor was non-zero, there would be a new SDT tensor at order $[2n+1]$ but such a tensor does not exist.  We thus see that  SD tensors, which are not traceless, have a zero symmetrized curl apart from the  only linearly independent exception (\ref{kaabb}). This ends the classification.

\end{document}